\documentclass[12pt]{iopart}
\usepackage{graphicx}
\usepackage{cite}	
\begin{document}










\title[Ne-like 3C and 3D lines under plasma environment]{Variation of the transition energies and oscillator strengths for the 3C and 3D lines of the Ne-like ions under plasma environment}

\author[C. S. Wu et al]{
Chensheng Wu,$^{1,2}$
Shaomin Chen,$^{1}$
T. N. Chang,$^{3}$
and Xiang Gao,$^{4,2}$
\\
$^{1}$Department of Engineering Physics, Tsinghua University, Beijing 100084, China\\
$^{2}$Beijing Computational Science Research Center, Beijing 100193, China\\
$^{3}$Department of Physics, Univ. of Southern California, LA, CA 90089-0484, U.S.A\\
$^{4}$Center for Computational Sciences, University of Tsukuba, 1-1-1 Tennodai, Tsukuba, Ibaraki 305-8577, Japan
}
E-mail: xgao@csrc.ac.cn

\date{Accepted XXX. Received YYY; in original form ZZZ}

\begin{abstract}
We present the results of a detailed theoretical study which meets the spatial and temporal criteria of the Debye-Hu\"ckel (DH) approximation on the variation of the transition energies as  well as the oscillator strengths for the ${2p^53d\ ^1P_1\rightarrow2p^6\ ^1S_0}$ (3C line) and the ${2p^53d\ ^3D_1\rightarrow2p^6\ ^1S_0}$ (3D line) transitions of the Ne-like ions subject to external plasma.\ Our study shows that the redshifts of the transition energy follow the general scaling behaviors similar to the ones for the simple H-like and He-like ions.\ Whereas the oscillator strength for the 3C line decreases, the oscillator strength for the spin-flipped 3D line increases as the strength of the outside plasma increases.\ As a result, their ratio is amplified subject to outside plasma environment.\ We further demonstrate that the plasma-induced variation between the relative strength of the 3C and 3D transitions is mainly due to the spin-dependent interactions which dictate the mixing of the $^1P_1$ component in the $^3D_1$ upper state of the 3D transition.\ In addition, we are able to find that the ratio between the relative oscillator strengths of the 3C and 3D lines in the presence of the plasma to their respective plasma-free values varies as a nearly universal function of $[(Z-9.2)DZ]^{-1.8}$, with $Z$ the nuclear charge and $D$ the Debye length.\ The results of this study should be of great help in the modelling and diagnostic of astrophysical plasmas as well as laboratory plasmas. \
\end{abstract}




\section{Introduction}
\indent The atomic and molecular data are indispensable physical parameters in astrophysics\ \cite{Tennyson2005,Kallman2007} and energy related researches\ \cite{Lindl2004,BooZer2004}.\ In particular, for the atomic ions in various astrophysical environment and laboratory experiments, it is important to understand the influence of the outside plasma to the atomic data as most of the ions actually existed in plasma environment.\ Data on atomic energy levels, oscillator strengths, and the related collision rates are required to simulate numerically the temporal-spatial motions for plasmas in astrophysical objects or in controlled fusion facilities\ \cite{Tennyson2005,Kallman2007,Lindl2004,BooZer2004} and to perform diagnostic analysis of the plasma parameters\ \cite{Griem1997,Skupsky1980}.\ Among astrophysical spectra, the X-ray emission lines corresponding to the ${2p^53d\ ^1P_1\rightarrow2p^6\ ^1S_0}$ (3C line) and the ${2p^53d\ ^3D_1\rightarrow2p^6\ ^1S_0}$ (3D line) transitions of the highly charged $\mathrm{Fe^{16+}}$ ions near $812$ eV are known to be two of the most prominent diagnostic lines\ \cite{Behar2001,Xu2002,Paerels2003,Trigo2013}.\ However, their diagnostic utility has been limited by the fact that although extensive studies have been carried out, discrepancies persist between the theoretical estimates and the measurements from the astrophysical and laboratory sources of the 3C/3D oscillator strengths ratio\ \cite{Brown2001a,Safronova2001,Chen2002,Chen2007,Bernitt2012,Natalia2014,Loch2015,Mendoza2017,Wang2017}.\ Although many studies were carried out recently to identify the reasons for such discrepancies, it remains as an open question\ \cite{Bernitt2012,Natalia2014,Loch2015,Mendoza2017,Wang2017}.\ Meanwhile, a recent study has demonstrated a substantial variation of the oscillator strengths for the $\alpha$-line of the H-like and He-like ions subject to outside dense plasmas\ \cite{Fang2018}.\ One of the main objectives of this work is to find out if such variation in the oscillator strengths, in particular the 3C/3D intensity ratio of the Ne-like ions, could also be shown in dense plasma environment.\\
\indent The influence of the plasma environment to the atomic processes is a challenging many-body problem and several sophisticated and complicated models were developed over the years to understand such processes\ \cite{Nguyen1986,Davis1982,Glenzer2009,Li2017}.\ On the other hand, the Debye-H\"uckel (DH) model, with its simplicity\ \cite{Debye1923}, could easily capture the important qualitative features over several orders of magnitude in plasma temperature and density, and, at the same time, offer the physical interpretation as a reliable reference for other theoretical models\ \cite{Chang2013,Chang2015,Fang2017,Fang2018,Bielinska2004,Ho_series,Kar2015,Saha2003,Deprince2017,Mukherjee_series,Zhang2010,Qi2009,Wang1995,Dai2001,Lopez1997,Okutsu2005,Shukla2011,Mondal2013}.\ Recently, we present a critical assessment of the DH approximation in terms of the spatial and temporal criteria for dense plasma which limits its applicability to atomic processes that are short-ranged in nature for the H-like and He-like ions with nucleus charge $Z$ between $5$ and $18$\ \cite{Chang2015,Fang2017,Fang2018}.\ This limited DH approximation has generated the redshift of the Lyman-$\alpha$ line of the H-like ion in a plasma environment in agreement both with the experimentally observed value and the data from more elaborate simulations based on quantum mechanical approaches\ \cite{Chang2015}.\ Its applications to the $\alpha$ emission lines of both hydrogen and helium-like ions which meet the spatial and temporal criteria have also led to simple scaling properties of the redshifts of the transition energies as well as the oscillator strengths\ \cite{Chang2015,Fang2017,Fang2018}.\ For more complex ions, it is expected that a full-relativistic application of the DH approximation is necessary to include the electron correlations as well as the relativistic effects.\ The main purpose of the present study is to explore the effect of the outside plasma environment to the change of the transition energy and the oscillator strength of the spectroscopically isolated lines due to the transitions of the ${3d}$ electron of the Ne-like ions based on the DH approximation.\\
\indent In section 2, we outline the theoretical procedures of our full relativistic calculations.\ Section 3 presents the results of our study and how the transition energies and oscillator strengths vary as the functions of Debye length in terms of the ratio of the temperature and the electron density of the outside plasma.\ Finally, in section 4, the implications of the present work are summarized.\\
\section{Theoretical Methods}
\indent Following the DH approximation, the potential for an atomic electron in plasma at a distance $r$ from a nuclear charge $Z$ is given by\ \cite{Chang2013,Chang2015,Rouse1967,Margenau1959},
\begin{equation}
V_\mathrm{DH}=\left\{
\begin{array}{ll}
V_i=-Z(\frac{1}{r}-\frac{1}{D+A}), & r\le A\\
\quad&\quad\\
V_o=-Z(\frac{De^{A/D}}{D+A})\frac{e^{-r/D}}{r}, & r> A.
\end{array} \right.
\label{equ1}
\end{equation}
where $A$ is the radius of the Debye sphere and the Debye length $D$ is expressed in terms of the Bohr radius $a_{0}$ by\ \cite{FFChen2006}
\begin{equation}
D=1.4048\times \sqrt{\frac{k_{b}T_{e}}{N_{e}}},
\label{equ2}
\end{equation}
where $N_{e}$ and $k_{b}T_{e}$ are the density and temperature of the free plasma electrons in the units of $10^{22}\ \mathrm{cm^{-3}}$ and $\mathrm{eV}$, respectively.\ We note that the atomic units are used throughout this paper if not otherwise specified.\\
\indent The application of the DH model depends on two key parameters\ \cite{Chang2013,Chang2015,Rouse1967,Margenau1959}.\ The first one is the radius $A$ of the Debye sphere, which separates the affected outside plasma environment and the slightly modified close-in region where the atomic characteristic dominates.\ The second one is the Debye length $D$, which is related to the electron density $N_e$ and temperature $T_e$ of the outside plasma based on the classical Maxwell-Boltzmann statistics.\ For the DH model and the classical Maxwell-Boltzmann statistics to apply for the atomic process in plasma environment, such processes should be short-ranged spatially, i.e., generally be limited to the  transitions involving the electron in the lower atomic states such as the present 3C and 3D lines.\ Temporally, the time scale characteristic of the atomic transition (i.e., the average lifetime of the upper state of the transition and/or the motion of the electron in its atomic orbital) should be considerably different (either longer or shorter) from the correlation time or the inverse of the plasma frequency $f_p=8.977\times 10^{3}N_{e}^{1/2}\ \mathrm{Hz}$ of the outside plasma\ \cite{FFChen2006}.\ More details in determining the DH parameters for the present calculation will be given in our subsequent discussion.\\
\indent For the relatively simple atomic systems, such as the H-like and He-like ions with relatively small $Z$ in our recent studies\ \cite{Chang2015,Fang2017,Fang2018}, the relativistic effect is negligible and the non-relativistic calculations should be adequate.\ But, in a more complex system with more electrons such as the Ne-like ions in the present calculation, the relativistic effects become important and should be taken into account.\ For the relativistic calculation, the N-electron Hamiltonian $\hat{H}_\mathrm{DH}$ for an atom in plasma environment is expressed as,
\begin{equation}
\begin{array}{c}
\hat{H}_\mathrm{DH}=\hat{H}_\mathrm{DC}+{\sum\limits_{i=1}^N} V_d(r_i,D),\\
\hat{H}_\mathrm{DC}={\sum\limits_{i=1}^N} c \vec{\alpha}\cdot\vec{p}_i+(\beta-1)c^2-\frac{Z}{r_i}+\sum\limits_{i<j}\frac{1}{r_{ij}},\\
V_d(r_i,D)=\frac{Z}{r_i}+V_\mathrm{DH},
\label{equ3}
\end{array}
\end{equation}
where $\hat{H}_\mathrm{DC}$ is the the Dirac-Coulomb Hamiltonian, $c$ is the speed of light, 
\begin{equation}
\hat{H}_\mathrm{DC}|\Gamma PJM\rangle_0=E^\mathrm{DC}_\Gamma |\Gamma PJM\rangle_0.
\label{equ4}
\end{equation}
where $P$ is the parity and $J$ and $M$ are the total angular momentum and magnetic quantum number, respectively.\ $\Gamma$ denotes the $\Gamma$th atomic eigenstate function.\ The ASFs are the N-electron eigenstate wave functions, which are the linear combinations of the configuration state functions (CSFs) with the same $P$, $J$ and $M$, namely,
\begin{equation}
|\Gamma PJM\rangle=\sum_{i=1}^{n_e} C_{i}^\Gamma|\gamma_i PJM\rangle,
\label{equ5}
\end{equation}
where $C_{i}^\Gamma$ is the expansion coefficient and $\gamma_i$ represents all other information to define the CSF uniquely.\ The CSFs, $|\gamma_i PJM\rangle$, which form a basis set for an $N$-electron atomic system in Hilbert space, are linear combinations of the Slater determinants of the atomic orbital wave functions (AOs).\ By applying the variational method to solve Equation (\ref{equ4}), one can obtain the mixing coefficients as well as the AOs self-consistently.\ This is known as the multi configuration self-consistent field method (MCSCF)\ \cite{Grant2006,Jonsson2007}.\\
\indent Our calculations were carried out using a revised multi-configuration Dirac-Fock (MCDF) approach in order to take the electron correlations into account adequately\ \cite{Han2012,Gao2014,Han2014,Gao2016}.\ The quasi-complete basis scheme is adopted to optimize the atomic orbitals (AOs) using the GRASP-JT version\ \cite{Han2012,Gao2014,Han2014,Gao2016} based on the earlier GRASP2K codes\ \cite{Jonsson2007}.\ In this calculation, AOs for ground and excited states are optimized separately.\ For the ground state, we optimize the spectroscopic AOs with the principal quantum number $n=1$, $2$, $3$ with $n-l-1$ nodes by MCSCF iterations to minimize the lowest $37$ energy levels of ${2p^6}$, ${2p^53s}$, ${2p^53p}$, and ${2p^53d}$ states to form the zeroth level basis.\ With the zeroth level basis fixed, the pseudo AOs with $n=4$ are optimized with further MCSCF iterations to minimize the ground state of the Ne-like ions to form the first level basis.\ The additional CSFs adopted in the optimization are generated by all single, double, and some triple excitations from the ${1s^22s^22p^6}$ and ${1s^22s^22p^53p}$ reference configurations to AOs with $n=3,\ 4$ (i.e., two electrons excited from the core shell and one electron excited from the valence shell).\ In succession, by adding more and more AOs in what we termed as the quasi-complete basis, we have included in the present calculation the quasi-complete basis with $n_\mathrm{max}=7$.\ For the excited state, the optimization processes are similar, except only $36$ energy levels of ${2p^53s}$, ${2p^53p}$, and ${2p^53d}$ states without the ground state are optimized by MCSCF iterations to obtain the zeroth level basis.\ The pseudo AOs are obtained in succession by further MCSCF iterations to optimize the first five $J^P=1^-$ excited states, with additional CSFs generated from the ${2p^53s}$ and ${2p^53d}$ reference configurations.\\
\indent In the second step, the plasma effect is included and calculated by the configuration interaction (CI) method based on these optimized AOs, i.e.,
\begin{equation}
\hat{H}_\mathrm{DH}|\Gamma PJM\rangle=E^\mathrm{DH}_\Gamma |\Gamma PJM\rangle.
\label{equ6}
\end{equation}
With the calculated ASFs, the oscillator strength of transition between atomic states $i$ and $j$ can be calculated as,
\begin{equation}
g_i f_{ij}=\frac{c\pi}{\omega^2}\sum_{M_i,M_j}|\langle \Gamma_j P_j J_j M_j|T^{(k)}_\lambda|\Gamma_i P_i J_i M_i\rangle|^2.
\label{equ7}
\end{equation}
where $T^{(k)}_\lambda$ is the radiative multipole operator\ \cite{Grant2006}, $\omega$ is the transition energy and $g_i=2J_i+1$ is the statistical weight of the initial state.\\
\indent It should be mentioned that we can solve the $\hat{H}_\mathrm{DH}$ in Equation (\ref{equ6}) directly by the MCSCF method where both the AOs and the expansion coefficients $C_{i}^\Gamma$ are varied with the Debye length $D$ and the radius of the Debye sphere $A$. When the CSFs are large enough, the results of this one-step method should be equivalent to our adopted two-step calculation procedure.\ It turns out, as we shall discussed in detail in Sec. 3, that the two-step calculation starting with plasma-free AOs in the first step is more convenient for a detailed analysis of the variation of the oscillator strength due to the influence of the outside plasma on the spin-dependent interactions for the 3C and 3D transitions of the Ne-like ions.\\
\section{Results and Discussions}
\begin{figure*}
\includegraphics[scale=0.5]{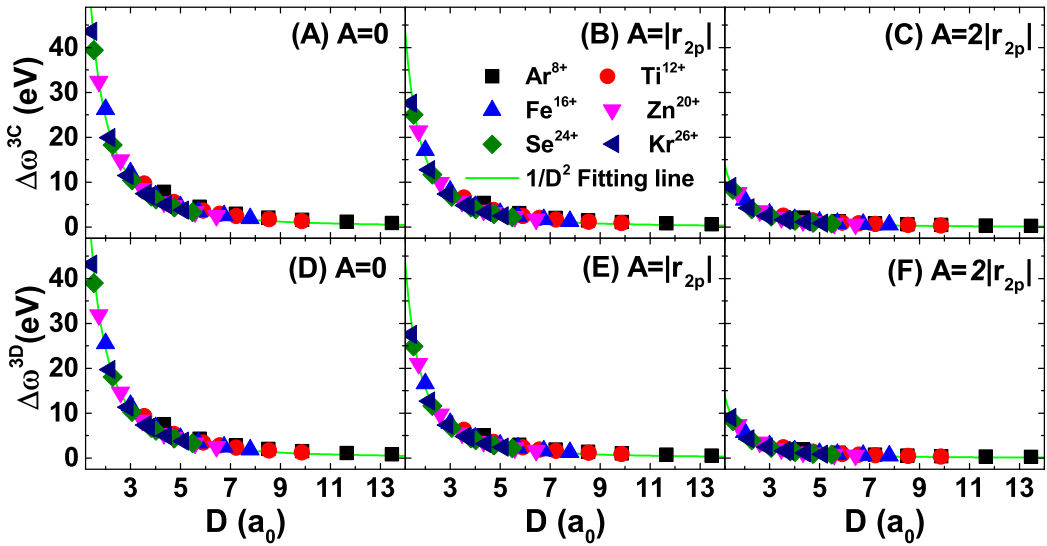}
\caption{The redshifts $\Delta_\omega^\mathrm{3C}$ and $\Delta_\omega^\mathrm{3D}$ in terms of $D$ with $A=0$, $A=|r_{2p}|$, and $A=2|r_{2p}|$.\ The green lines represent the $1/D^2$ fitting.}
\label{fig2}
\end{figure*}

\begin{figure*}
\includegraphics[scale=0.5]{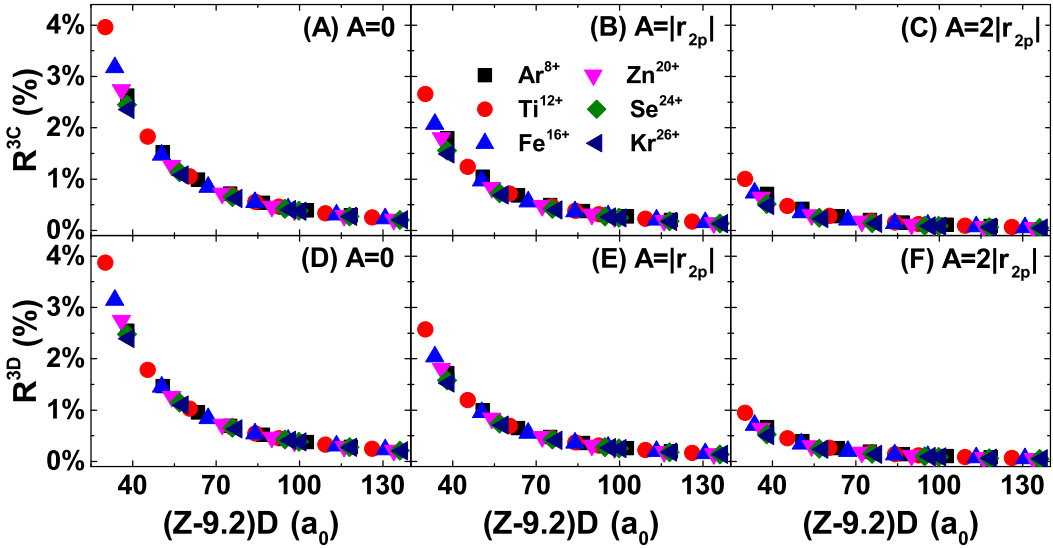}
\caption{The ratios $R^\mathrm{3C}$ and $R^\mathrm{3D}$ between the redshift and the plasma-free energy of the 3C and 3D lines as functions of the reduced Debye length $\lambda_D=(Z-9.2)D$ with $A=0$, $A=|r_{2p}|$, and $A=2|r_{2p}|$. }
\label{fig3}
\end{figure*}

\begin{figure}
\includegraphics[scale=0.3]{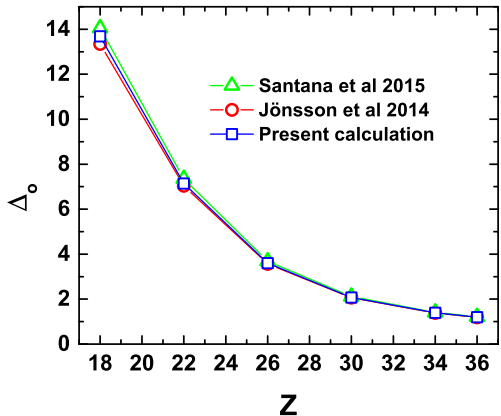}
\caption{The comparison between the present results and the ones from other earlier calculations for the ratios $\Delta_0= gf_0^{3C}/gf_0^{3D}$ of the plasma-free oscillator strengths between the 3C and 3D lines. }	
\label{fig4}
\end{figure}

\begin{figure*}
\includegraphics[scale=0.5]{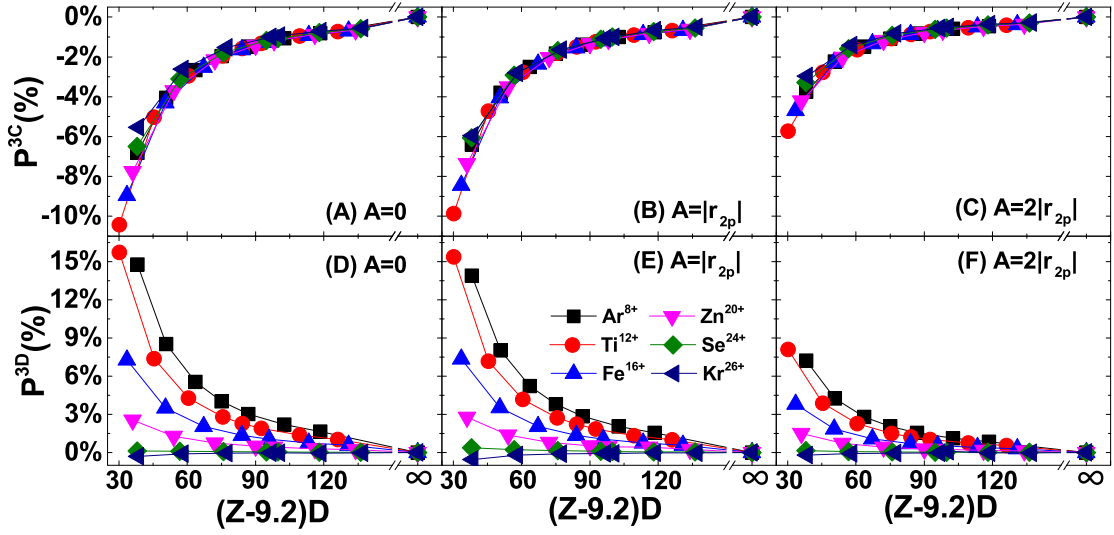}
\caption{The percentage change $P^\mathrm{3C}$ and $P^\mathrm{3D}$ of the oscillator strengths of the 3C and 3D lines as the functions of the reduced Debye length $\lambda_D=(Z-9.2)D$ with $A=0$, $A=|r_{2p}|$, and $A=2|r_{2p}|$.}
\label{fig5}
\end{figure*}

\indent Most of the existing plasma measurements are carried out with the electron densities from $10^{18}$ to $10^{24}\ \mathrm{cm^{-3}}$ and electron temperature $k_{b}T$ from a few $\mathrm{eV}$ to hundreds of $\mathrm{eV}$\ \cite{Nantel1998,Saemann1999,Leng1995,Woolsey1998,Goldston1995}.\ The plasma frequency $f_p=8.977\times 10^{3}N_{e}^{1/2}\ \mathrm{Hz}$\ \cite{FFChen2006} corresponding to such densities range from about $10^{13}$ to $10^{16}\ \mathrm{Hz}$ or with the period $\tau _{p}=1/f_{p}$ in the range of $10^{-13}\sim10^{-16}\ \mathrm{sec}$.\ To apply the DH approximation to Ne-like ions subject to outside plasma for transitions involving ${2p}$ and ${3d}$ electrons, we started our study by first identifying the range of Z that meets the criteria discussed in Sec. 2.\\
\indent The characteristic time of the motion of ${2p}$ and ${3d}$ electrons for hydrogen atom are about $1.2\times 10^{-15}$ and $4.1\times 10^{-15}\ \mathrm{sec}$, respectively\ \cite{Chang2015}.\ For our interested Ne-like ions with an effective charge $Z_{eff}$, the characteristic time could be estimated by using the $1/Z_{eff}^2$ scaling relation, similar to the hydrogen-like ions.\ Starting from $\mathrm{Ar^{8+}}$ ion with an estimated $Z_\mathrm{eff}=Z-9$, the characteristic time of ${2p}$ and ${3d}$ electrons would be $(1.2\times 10^{-15}/9^2)=1.5\times10^{-17}$ and $(4.1\times 10^{-15}/9^2)=5.1\times10^{-17}\ \mathrm{sec}$, respectively, which are sufficiently smaller than the period $\tau _{p}$ estimated above for plasma with $N_e$ less than $10^{24}\ \mathrm{cm^{-3}}$.\ This implies that the DH approximation would work for the Ne-like ions with $Z$ greater than 18.\ On the other hand, for DH approximation to apply, the transition rates should also be different from the $f_{p}$ given earlier\ \cite{Chang2015}.\ It turns out that the transition rate of $\mathrm{Ar^{8+}}$ is of the order of $10^{12}\ \mathrm{sec^{-1}}$ and increases to about $10^{14}\ \mathrm{sec^{-1}}$ for Ne-like $\mathrm{Kr^{26+}}$, which corresponds to a density $N_e$ about $10^{21}\ \mathrm{cm^{-3}}$.\ For Ne-like ions with $Z$ greater than 37, the transition rate would be greater than $10^{14}\ \mathrm{sec^{-1}}$ and overlap with the plasma frequency of $10^{14}\ \mathrm{Hz}$, thus leading to the breakdown of the DH approximation.\ As a result, the DH approximation might work for the 3C and 3D transitions of the Ne-like ions with $Z$ between 18 and 36 for dense plasma with density $N_e$ between $10^{21}\ \mathrm{cm^{-3}}$ and $10^{24}\ \mathrm{cm^{-3}}$.\ Based on the same temporal criterion, the DH approximation should also work for all Neon-like ions with $Z$ greater than 18 at a plasma density less than $10^{16}\ \mathrm{cm^{-3}}$.\ However, at such density, the Debye length $D$ would be several orders of magnitude greater than the size of the ions and the atomic transition is not expected to be significantly influenced by the outside plasma environment.\\
\indent The DH approximation also requires that the Debye length be greater than the radius of the Debye sphere $A$\ \cite{Chang2013,Chang2015}.\ Similar to our previous study for the H-like and He-like ions, in the present study, we have chosen $A$ to be expressed approximately in terms of the average of the ${2p}$ orbit, or, its order of magnitude of $|r_{2p}|$ with $|r_{2p}|=2^2/Z_\mathrm{eff}\ a_0$.\ For comparison, we have included in our calculation with $A=0$, $|r_{2p}|$ and $2|r_{2p}|$.\ The smallest $D$ in the present calculation ranges from $4.33\ a_0$ to $1.44\ a_0$ for $\mathrm{Ar^{8+}}$ to $\mathrm{Kr^{26+}}$ ions, respectively, which are still substantially greater than the average radius of the $n=3$ orbitals.\\
\indent With the electron correlations adequately taken into account using the quasi-complete basis presented in Sec. 2, Table\ \ref{tab1} shows that the plasma-free transition energies of the 3C and 3D lines from the present calculation are in good agreement with the experimental values\ \cite{Brown2001,Santana2015,NIST,Beiersdorfer2001} to $0.1\%$ or less.\ The redshifts $\Delta\omega(D)$ of the transition energies of 3C and 3D in terms of $D$ are given by,
\begin{equation}
\begin{array}{cc}
\Delta\omega^\mathrm{3C}(D)=&[E_{2p^{6},\ ^1S_{0}}(D)-E_{2p^{6},\ ^1S_{0}}(\infty)]-\\
&[E_{2p^{5}3d,\ ^1P_{1}}(D)-E_{2p^{5}3d,\ ^1P_{1}}(\infty)],\\
\Delta\omega^\mathrm{3D}(D)=&[E_{2p^{6},\ ^1S_{0}}(D)-E_{2p^{6},\ ^1S_{0}}(\infty)]-\\
&[E_{2p^{5}3d,\ ^3D_{1}}(D)-E_{2p^{5}3d,\ ^3D_{1}}(\infty)].
\end{array}
\label{equ8}
\end{equation}
\noindent Qualitatively, in a single electron approximation, the redshifts $\Delta\omega^\mathrm{3C}(D)$ and $\Delta\omega^\mathrm{3D}(D)$ subject to external plasma could be estimated as the difference of the expectation values of $\langle 2p|\Delta V_D |2p\rangle$ and $\langle 3d|\Delta V_D |3d\rangle$, where $\Delta V_D=(Ze^2/r)(1-e^{-r/D})$ is the difference  between the pure Coulomb potential and the screening Coulomb potential.\ The redshift $\Delta\omega(D)$ is then given approximately by\ \cite{Bieli2004},
\begin{equation}
\Delta\omega(D)\sim [\frac{11}{4}-\frac{16}{ZD}+O(\frac{1}{(ZD)^2})]/D^2,
\label{equ9}
\end{equation}
or, to the first order approximation, inversely proportional to $D^2$, similarly to Equation (13) of\ \cite{Fang2017} for the H-like and He-like ions.\ Figure \ref{fig2} presents the variation of our calculated redshifts $\Delta\omega(D)$ of the 3C and 3D lines of different Ne-like ions that decrease as $1/D^2$ with increasing $D$.\ The slight deviation from the $1/D^2$ for the low $Z$ ions, such as the $\mathrm{Ar^{8+}}$ and $\mathrm{Ti^{12+}}$, is due primarily to their relatively stronger electron correlation.\\
\begin{table}
\caption{The plasma free transition energies of 3C and 3D lines of different $Z$ ions in comparison with the experimental values.}
\begin{indented}
\item[]\begin{tabular}{ccccc}
\br
\ &$\omega^{3C}$ (eV)&$\omega^{3C}_\mathrm{Exp}$&$\omega^{3D}$ (eV)&$\omega^{3D}_\mathrm{Exp}$\\
\hline
Ar&$298.950$&$298.894^{\rm a}$&$295.158$&$295.159^{\rm a}$\\
Ti&$530.984$&$530.850^{\rm b}$&$523.273$&$523.190^{\rm b}$\\
Fe&$825.853$&$825.772^{\rm c}$&$812.390$&$812.406^{\rm c}$\\
Zn&$1185.443$&$1185.143^{\rm c}$&$1162.564$&$1162.202^{\rm c}$\\
Se&$1612.066$&$1611.193^{\rm d}$&$1573.802$&$1573.367^{\rm d}$\\
Kr&$1851.318$&$1850.976^{\rm c}$&$1802.361$&$1802.175^{\rm c}$\\
\br
\end{tabular}
\item[]{$^{\rm a}$The results from Santana et al (2015)\ \cite{Santana2015}.}
\item[]{$^{\rm b}$The results from NIST (2018)\ \cite{NIST}.}
\item[]{$^{\rm c}$The results from Brown et al (2001)\ \cite{Brown2001}.}
\item[]{$^{\rm d}$The results of Beiersdorfer et al (2001)\ \cite{Beiersdorfer2001}.}
\label{tab1}
\end{indented}
\end{table}
\indent Our previous works on the H-like and He-like ions subject to outside plasma environment has led to a general feature that the ratio $R=\Delta\omega/\omega_0$ between the redshift $\Delta\omega$ and its plasma-free energy $\omega_0$ varies at the same rate in terms of a reduced Debye length $\lambda_D=Z_\mathrm{eff}D$.\ For the Ne-like ions in the current study, we found that the ratio $R$ is also a function of the reduced Debye length with $Z_\mathrm{eff}=Z-9.2$ for different Debye radius $A$ as shown in Figure\ \ref{fig3}.\ The small difference from $Z_\mathrm{eff}=Z-9$ reflects the slightly stronger electron correlation for Ne-like ions.\\
\indent Like most of the atomic transitions, the contribution to the plasma-free oscillator strength $gf^\mathrm{3C}_0$ of the 3C line is primarily due to the dipole transition of the upper $^1P_1$ state to the $^1S_0$ ground state of the Ne-like ions.\ On the other hand, for the spin-flipped 3D line, the contribution to its plasma-free oscillator strength $gf^\mathrm{3D}_0$ comes mainly from the dipole transition to the $^1S_0$ ground state from the limited $^1P_1$ component mixed in its upper $^3D_1$ state due to the spin-dependent interactions.\ With the electron correlation and the spin-dependent interactions carefully taken into account in the present calculation, Figure\ \ref{fig4} shows that our calculated ratios $\Delta_0=gf^\mathrm{3C}_0/gf^\mathrm{3D}_0$ of the plasma-free oscillator strengths between the 3C and 3D lines are in close agreement with other elaborated theoretical calculations for Ne-like ions with $Z$ between 18 and 36\ \cite{jonsson2014,Santana2015}.\ Whereas the upper states of the 3C line for all Ne-like ions are dominated by the $^1P_1$ component, the mixing of the limited $^1P_1$ component for the small-$Z$ ion in the upper $^3D_1$ state of the 3D line due to the spin-dependent interactions is expected to be enhanced for the higher $Z$ ions.\ Although the oscillator strength $gf^\mathrm{3C}_0$ does not change very much as $Z$ increases, the oscillator strength $gf^\mathrm{3D}_0$ would increase at a rate noticeably more than that of the $gf^\mathrm{3C}_0$.\ Together with the fairly small plasma-free oscillator strength $gf^\mathrm{3D}_0$ for the ions of smaller $Z$, the ratio $\Delta_0$ is therefore relatively large for the Ne-like ion with smaller $Z$ in comparison to the ones from ions of larger $Z$ as shown in Figure\ \ref{fig4}.\\
\indent Similar to our earlier study of the H-like and He-like ions for the effect of the outside plasma to the oscillator strengths, we started by examining the relative oscillator strengths $gf_r(D)$ of the 3C and 3D lines in terms of the ratios of the oscillator strengths subject to external plasma to their respective plasma-free values $gf_0$, i.e., $gf^\mathrm{3C}_r(D)=gf^\mathrm{3C}(D)/gf^\mathrm{3C}_0$ and $gf^\mathrm{3D}_r(D)=gf^\mathrm{3D}(D)/gf^\mathrm{3D}_0$.\ Figure \ref{fig5} presents the percentage change of the oscillator strengths of the 3C and 3D line, i.e., $P^{3C}=gf^\mathrm{3C}_r(D)-1$ and $P^{3D}=gf^\mathrm{3D}_r(D)-1$ as functions of the reduced Debye length $\lambda_D=(Z-9.2)D$ with $A=0$, $A=|r_{2p}|$, and $A=2|r_{2p}|$.\ Qualitatively, the plasma effect on the atomic orbitals is stronger with smaller Debye radius $A$ when the influence of the outside plasma extended closer to the nucleus due to the less attractive $V_{DH}$ outside the Debye sphere.\ Indeed, our calculation has shown that the decrease of the percentage changes $P^{3C}$ with decreasing $\lambda_D$ is noticeably less for the ones shown in plots C and F of Figure\ \ref{fig5} with $A=2|r_{2p}|$ than the ones in plots A and D with $A=0$.\ This is due to the fact that with the larger $A=2|r_{2p}|$, the inner portion of the $n=3$ orbitals are effectively free from the plasma effect, whereas with $A=0$, the $n=3$ orbitals are subject entirely to the outside plasma.\\
\indent To further understand the variations of the oscillator strengths of 3C and 3D lines, it's convenient to express the oscillator strength in Equation (\ref{equ7}) under the dipole long wavelength approximation in length gauge as,
\begin{equation}
g_{\alpha}f_{\alpha\beta}\sim \omega_{\alpha\beta}\cdot|\sum_{i,j}C_{i}^{\Gamma_\alpha} C_j^{\Gamma_\beta}\langle \gamma_i P_i J_i M_i|\hat{r}|\gamma_j P_j J_j M_j\rangle|^2,
\label{equ10}
\end{equation}
where $\gamma_i$ and $\gamma_j$ denote different set of configuration state functions (CSFs) with their corresponding expansion coefficients $C_{i}^{\Gamma_{\alpha}}$ and $C_{j}^{\Gamma_{\beta}}$ for the initial and final states and $\langle \gamma_i P_i J_i M_i|\hat{r}|\gamma_j P_j J_j M_j\rangle$ the dipole matrix elements for the transition from state $i$ to state $j$.\ In the present study, the AOs are fixed in the first step for the plasma-free CI calculation.  The plasma effect will only affect the atomic state functions (ASFs) calculated from Equation (\ref{equ6}) and the oscillator strengths via their corresponding expansion coefficients $C_{i}^{\Gamma_{\alpha}}$ and $C_{j}^{\Gamma_{\beta}}$.\\
\indent Since the final state for both the 3C and 3D lines is the nearly pure spin-singlet $2p^6\ ^1S$ state, it is easier to examine these two transitions by applying the selection rule $\Delta S=0$ with the total spin as the good quantum number in the $LS$ coupling than the $jj$ coupling employed in our relativistic calculation.\ As a result, we will start our discussion by first transforming the expansion coefficients of the $jj$-coupled ASFs, calculated from Equation (\ref{equ6}) in the presence of external plasma, into the $LS$-coupled ones.\ Table \ref{tab2} presents the geometric transformation matrix between the three prominent $jj$-coupled and $LS$-coupled ASFs corresponding to the $2p^53d$ configuration of the upper $J^\pi = 1^-$ states of the 3C and 3D lines.\ From the $\Delta S=0$ selection rule, only the $2p^53d\ ^1P$ component of the upper states due to the mixing of the spin singlet and triplet states from the relativistic effects of the spin-dependent interactions will contribute to the oscillator strengths of the 3C and 3D lines.\ It turns out that our calculation has found that two of the three $2p^53d$ components, i.e., $^1P_1$ and $^3D_1$, contribute over $90\%$ of the upper states of the 3C and 3D lines.\ Hence, qualitatively, the upper states of the 3C and 3D lines could be further approximated by the two-state system with the expansion coefficients in terms of a parameter $\theta$ and expressed compactly as,
\begin{table*}
\caption{The geometric transformation matrix between the jj-coupling scheme and the LS-coupling scheme for $J^{\pi}=1^-$.}
\begin{tabular}{c|ccc}
\br
 &${[(2p_{1/2}^22p_{3/2}^3)_{3/2}3d_{3/2}]_{3/2}}$&${[(2p_{1/2}^22p_{3/2}^3)_{3/2}3d_{5/2}]_{3/2}}$&${[(2p_{1/2}^12p_{3/2}^4)_{1/2}3d_{3/2}]_{3/2}}$\\
\hline
${2p^53d, ^3P_1}$&$\sqrt{\frac{8}{15}}$&$-\sqrt{\frac{3}{10}}$&$\sqrt{\frac{1}{6}}$\\
${2p^53d, ^3D_1}$&$-\sqrt{\frac{2}{5}}$&$-\sqrt{\frac{1}{10}}$&$\sqrt{\frac{1}{2}}$\\
${2p^53d, ^1P_1}$&$\sqrt{\frac{1}{15}}$&$\sqrt{\frac{3}{5}}$&$\sqrt{\frac{1}{3}}$\\
\br
\end{tabular}
\label{tab2}
\end{table*}
\begin{equation}
\begin{array}{ll}
|\Psi^\mathrm{3C}\rangle=\cos\theta|{2p^53d,\ ^1P_1}\rangle+\sin\theta|{2p^53d,\ ^3D_1}\rangle,\\
|\Psi^\mathrm{3D}\rangle=\sin\theta|{2p^53d,\ ^1P_1}\rangle-\cos\theta|{2p^53d,\ ^3D_1}\rangle,
\end{array}
\label{equ11}
\end{equation}
where, $\sin\theta$ and $\cos\theta$ are the normalized expansion coefficients in the $LS$-coupled scheme.\ We should also point out that the effect of the spin-dependent interactions is due mostly to the one-particle spin-orbit interaction, i.e., $H_{SO}\sim (1/r)(\partial V_{DH}/\partial r)\hat{L}\cdot \hat{S}$ with the potential $V_{DH}$ given by Equation (\ref{equ1}).\ And, with a $Z/r^3$ dependence for $H_{SO}$, the region close to the nucleus, i.e., with $r$ much smaller than the Debye length $D$, dictates almost entirely the mixing of the $^1P_1$ and $^3D_1$ states in $\Psi^\mathrm{3C}$ and $\Psi^\mathrm{3D}$.\\
\begin{figure}
\includegraphics[scale=0.3]{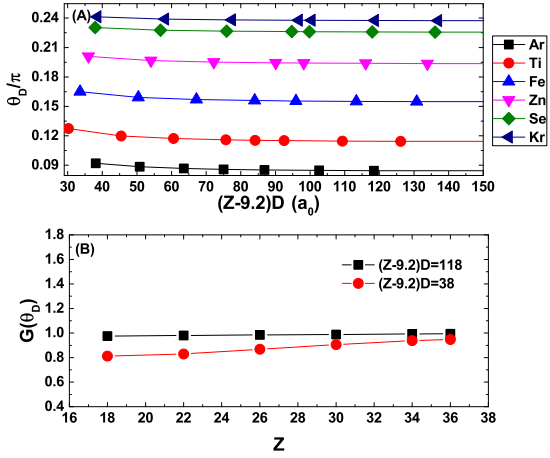}
\caption{The $\theta_D$ and $G(\theta_D)$ in the two-state approximation with $A=0$.}
\label{fig6}
\end{figure}
\indent The upper plot of Figure\ \ref{fig6} presents our calculated $\theta_D$ of the Ne-like ions with different $Z$ as functions of the reduced Debye length $\lambda_D=(Z-9.2)D$ with $A=0$.\ As expected, for the low-Z $\mathrm{Ar^{8+}}$ ion, the upper state $\Psi^\mathrm{3C}$ is dominated by its $^1P_1$ component, or, $\cos^2\theta_D$, with smaller $\theta_D$ due to the relatively weak spin-orbit interaction.\ For the higher-Z $\mathrm{Kr^{26+}}$ ion with stronger spin-orbit interaction, the $\theta_D$ is substantially greater.\ The parameter $\theta_D$ increases for all Ne-like ions as the effect of the external plasma increases, i.e., when $D$ decreases.\ Qualitatively, this could be understood by comparing the dominant spin-orbit interaction, which leads to the mixing of the $^1P_1$ and $^3D_1$ states, to the Debye potential $V_{DH}$, i.e., the ratio $H_{SO}/V_{DH}$.\ Near the nucleus, where $r$ is much smaller than $D$, $H_{SO}/V_{DH}$ is given approximately by $\sim (1/{Dr}+1/r^2)\hat{L}\cdot\hat{S}$.\ The enhancement of the ratio $H_{SO}/V_{DH}$, as $D$ decreases, will lead to an increase in $\theta$ as shown in Figure\ \ref{fig6} from our calculation.\ When $\lambda_D$ decreases, the $^1P_1$ component of $\Psi^\mathrm{3C}$ will decrease due to an increasing $\theta_D$, or, a decreasing $\cos^2\theta_D$.\ Since the rate of change for $\cos^2\theta_D$ is small with the relatively small $\theta_D$ for all Ne-like ions shown in Figure\ \ref{fig6}, the rate of decrease in $gf^\mathrm{3C}(D)$ from its plasma-free value of $gf^\mathrm{3C}_0$ for the 3C line is fairly small.\ On the other hand, when $\lambda_D$ decreases, the $^1P_1$ component of $\Psi^\mathrm{3D}$ will increase like $\sin^2\theta_D$ with increasing $\theta_D$, leading to a rate of increase in $gf^\mathrm{3D}(D)$ from its plasma-free value of $gf^\mathrm{3D}_0$ substantially greater than the one for the 3C line.\ The fact that $gf^\mathrm{3C}(D)$ and $gf^\mathrm{3D}(D)$ have inversed variations with $D$, its ratio $\Delta_I=gf^\mathrm{3C}(D)/gf^\mathrm{3D}(D)$, or alternatively, $\Delta_r=gf^\mathrm{3C}_r(D)/gf^\mathrm{3D}_r(D)=\Delta_I/\Delta_0$, would be more sensitive to the plasma environment.\ For example, Figure\ \ref{fig5} shows that at $\lambda_D=30$ with $A=0$ for $\mathrm{Ti^{12+}}$ ion, the ratio $gf^\mathrm{3C}_r(D)$ decreases by about $10\%$ and the ratio $gf^\mathrm{3D}_r(D)$ increases by about $16\%$, whereas the ratio $\Delta_r$ decreases by close to $25\%$.\\
\indent Based on our two-step procedure, it is straightforward to extend the theoretical oscillator strength calculation for the Ne-like ions from one to another.\ On the other hand, it is far more complicated to extend the measurement of the oscillator strength from one ion to the next experimentally.\ Our calculation has shown in Figure\ \ref{fig5} that the individual features of the variation of the oscillator strength for the 3C and 3D lines of the Ne-like ions vary in the opposite direction and do not follow a simple universal function for all ions as the reduced Debye length varies.\ For the diagnostic possibility based on the extrapolation to other ions from the known and confirmed theoretical calculation and the experiment observed data of one particular ion subject to outside plasma, it is necessary to identify a general feature or scaling relation involving the variation of the oscillator strengths that is similar to but different from the one we have identified for the H-like and He-like ions.\ As we already pointed out that the variation of the ratio $\Delta_r=gf^\mathrm{3C}_r(D)/gf^\mathrm{3D}_r(D)$ is more sensitive to the change in $D$, we will now focus our discussion on this ratio to find out if a simple general feature could be identified due to the change of external plasma environment.\\
\indent Since the dipole matrix element in Equation (\ref{equ10}) in our two-step calculation is independent of the external plasma, the ratio $\Delta_r$ could be expressed as,
\begin{equation}
\frac{gf^\mathrm{3C}_r}{gf^\mathrm{3D}_r}=\frac{\omega^\mathrm{3C}(\lambda_D)}{\omega^\mathrm{3D}(\lambda_D)}\cdot\frac{\omega^\mathrm{3D}_0}{\omega^\mathrm{3C}_0}\cdot\frac{\cos^2{\theta_D}}{\sin^2{\theta_D}}\cdot\frac{\sin^2{\theta_0}}{\cos^2{\theta_0}}.
\label{equ12}
\end{equation}
The ratio $\omega^\mathrm{i}(\lambda_D)/\omega^\mathrm{i}_0$ on the right-hand side of Equation (\ref{equ12}) equals $[\omega^\mathrm{i}_0-\Delta\omega^\mathrm{i}(\lambda_D)]/\omega^\mathrm{i}_0=1-R^\mathrm{i}(\lambda_D)$, where $R^\mathrm{i}(\lambda_D)$, or, $R^\mathrm{3C}$ and $R^\mathrm{3D}$, are the nearly universal functions of $\lambda_D$ shown in Figure\ \ref{fig3}.\ Equation (\ref{equ12}) would then take the form of,
\begin{equation}
\Delta_r = [\frac{1-R^\mathrm{3C}(\lambda_D)}{1-R^\mathrm{3D}(\lambda_D)}]G(\theta_D),
\label{equ13}
\end{equation}
where
\begin{equation}
G(\theta_D)=\frac{\cos^2\theta_D/\cos^2\theta_0}{\sin^2\theta_D/\sin^2\theta_0},
\label{equ14}
\end{equation}
represents the relative contributions from the $^1P_1$ component of the $\Psi^\mathrm{3C}$ and $\Psi^\mathrm{3D}$ to the variation of the oscillator strength due to the spin-dependent interactions.\ The bottom plot of Figure\ \ref{fig6} shows approximately that $G(\theta_D)$ varies linearly to the nuclear charge $Z$.\ As a result, the ratio $\Delta_r$ could be expressed as a nearly universal function in terms of $\lambda_DZ$.\ For experimental measurement, it is more convenient to express the plasma induced variation of the oscillator strengths in terms of the percentage reduction $\Delta_p$ of the ratio $\Delta_I$ from the plasma-free $\Delta_0$, or, $\Delta_p=(\Delta_0-\Delta_I)/\Delta_0=1-\Delta_r$.\ Figure\ \ref{fig7} shows that our calculated $\Delta_p$ with $A=0$, $|r_{2p}|$, and $2|r_{2p}|$ could all be expressed approximately in terms of nearly universal functions following closely to $[(Z-9.2)ZD]^{-1.8}$ for all Ne-like ions.\ This general feature offers the possibility to extrapolate the calculated data from one particular Ne-like ion to other Ne-like ions in plasma environment based on the application of DH approximation.\\
\begin{figure}
\includegraphics[scale=0.3]{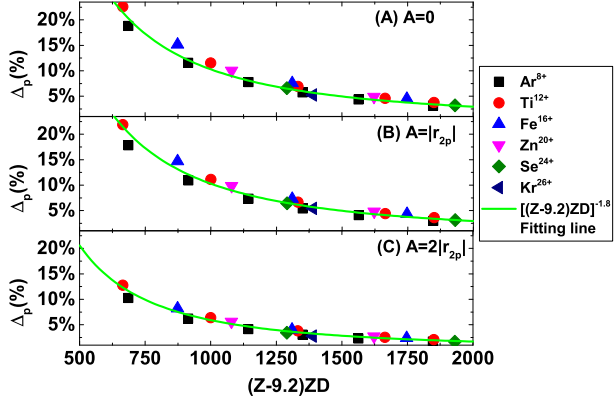}
\caption{The percentage reduction $\Delta_p$ of the intensity ratio in terms of $(Z-9.2)ZD$ with $A=0$, $A=|r_{2p}|$, and $A=2|r_{2p}|$. }
\label{fig7}
\end{figure}
\indent From our calculated results presented in Figure\ \ref{fig7}, we find that for the Ne-like $\mathrm{Fe^{16+}}$ ions, shown in the top plot of Figure\ \ref{fig8}, the ratio $\Delta_I=gf^\mathrm{3C}(D)/gf^\mathrm{3D}(D)$ would indeed decrease substantially as $D$ decreases and result in a ratio close to the measured value from the coulped X-ray free electron laser (XFEL) and the electron beam ion trap (EBIT) experiment performed at the Linear Coherent Light Source (LCLS)\ \cite{Bernitt2012}.\ However, it turns out that such decrease in $\Delta_I$, based on our estimate, would only occur if the outside plasma density is several orders of magnitude higher than those of the LCLS experiment according to the bottom plot of Figure\ \ref{fig8}.\ As a result, the outside plasma could not be the cause that is responsible for the disagreement between the theory and LCLS experiment.\ In contrast, the scale of the astrophysical plasma conditions can vary several orders of magnitude\ \cite{Goldston1995}, the present results may still have great significance in the modelling and diagnostic of astrophysical plasmas.\\
\indent We note in a recent theoretical study\ \cite{Mendoza2017}, by manipulating the relativistic wave functions with a set of \textit{a priori} scaling parameters, a reduction in the intensity ratio $\Delta_I=gf^\mathrm{3C}/gf^\mathrm{3D}$ close to the measured value for Ne-like $\mathrm{Fe^{16+}}$ ions was derived.\ Unlike the plasma-free transition energies presented in Table\ \ref{tab1} from the present calculation, which are less than $0.1\%$ from the observed data, some of the fitted low-lying energies in\ \cite{Mendoza2017} are different from the known measured energy levels by as high as $7-8\%$.\ It is possible that by varying the scaling parameters, the effect on the wave functions may have mimicked the spin-dependent interactions discussed earlier and thus leading to the ratio $\Delta_I$ close to the experimental result.\ Since the recent measured ratio between the intensity of the 3C to 3D emission lines was performed under the experimental condition with a substantially lower plasma density than those investigated in the present study as we pointed out earlier, the difference between the atomic structure calculation and the measured data might be due to other atomic processes, beyond an accurate characterization of the electron-electron correlation and the spin-dependent interactions.\ One such possibility could be an increase in the upper state population of the 3D line due to other atomic process such as the resonance induced population transfer from the nearby Na-like $\mathrm{Fe^{15+}}$ ions with the same $Z$, even when the outside plasma density is substantially smaller\ \cite{Wu2019}.\\
\begin{figure}
\includegraphics[scale=0.3]{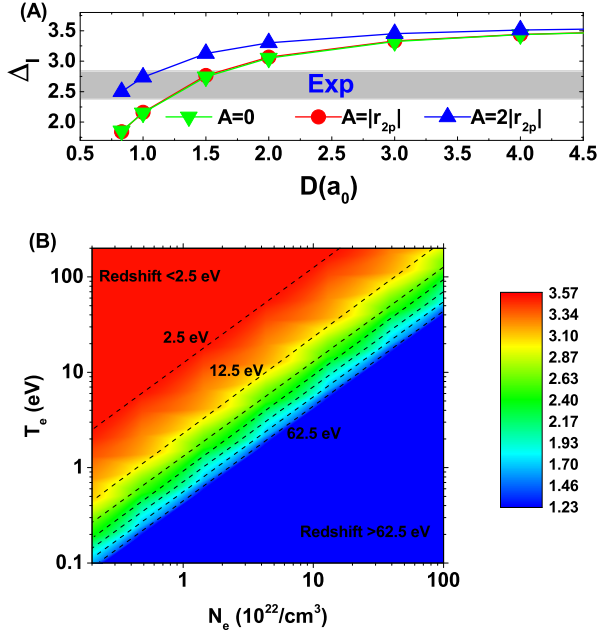}
\caption{The intensity ratio $\Delta_I=gf^\mathrm{3C}/gf^\mathrm{3D}$ decreases as $D$ varies from $4$ $a_0$ to $1$ $a_0$ for the Ne-like $\mathrm{Fe^{16+}}$ ion.\ The red zone in \ref{fig8}(B) represents those with $\Delta_I$ greater than $3.53$ and the blue zone with smaller than $1.23$.\ The experimental results are in the green zone. }
\label{fig8}
\end{figure}
\section{Conclusion}
\indent Following the spatial and temporal criteria for the DH model, we present a study on the 3C and 3D emission lines for the Ne-like ions with nuclear charge $Z$ between 18 and 36 subject to outside plasma environment.\ First, similar to the $\alpha$-emission line of the H-like and He-like ions subject to outside plasma, our relativistic calculation has shown that the redshift of the 3C and 3D emission lines of all Ne-like ions, i.e., $\Delta\omega^\mathrm{3C}$ and $\Delta\omega^\mathrm{3D}$, increases as the Debye length $D$ decreases.\ Qualitatively, we have shown that both $\Delta\omega^\mathrm{3C}$ and $\Delta\omega^\mathrm{3D}$ vary approximately at a rate of $1/D^2$.\ In addition, we have demonstrated that the ratio of the redshifts to the plasma-free energy $\omega_0$ of the 3C or 3D lines, i.e., $R^\mathrm{3C}=\Delta\omega^\mathrm{3C}/\omega^\mathrm{3C}_0$ or $R^\mathrm{3D}=\Delta\omega^\mathrm{3D}/\omega^\mathrm{3D}_0$, for different Ne-like ions follows a nearly universal function of the reduced Debye length $\lambda_D=(Z-9.2)D$.\\
\indent Like most of the atomic transitions, the contribution to the plasma-free $gf^\mathrm{3C}_0$ of the 3C line is primarily due to the dipole transition of the upper $^1P_1$ state to the $^1S_0$ ground state of the Ne-like ions.\ Similar to the $\alpha$-emission line of the H-like and He-like ions, the oscillator strengths $gf^\mathrm{3C}(D)$ of the 3C line of the Ne-like ions of different $Z$ decrease with similar ratios $gf^\mathrm{3C}_r=gf^\mathrm{3C}(D)/gf^\mathrm{3C}_0$ when the Debye length $D$ decreases, or, when the effect of the outside plasma increases.\ In contrast, for the 3D line, the variation of the oscillator strength ratio $gf^\mathrm{3D}_r=gf^\mathrm{3D}(D)/gf^\mathrm{3D}_0$ of the Ne-like ion increases as the effect of the outside plasma increases.\ Unlike the 3C line, the variations of $gf^\mathrm{3D}_r$ are noticeably different for ions of different $Z$, with larger variations for ions of smaller $Z$.\ This is in part due to the relatively smaller $^1P_1$ component of the upper $^3D_1$ state due to the weaker spin-dependent interactions and consequently the smaller plasma-free oscillator strength $gf^\mathrm{3D}_0$ for the ions of smaller $Z$ than that for the ions of larger $Z$.\ In addition to the smaller $gf^\mathrm{3D}_0$, the proportionally larger increase of the $^1P_1$ component mixed in its upper $^3D_1$ state due to the spin-dependent interaction subject to outside plasma further enhances the variation of the ratio $gf^\mathrm{3D}_r$ for the ions of smaller $Z$.\\
\indent Although the calculated $gf^\mathrm{3D}_r$ values for different Ne-like ions are very different for ions of different $Z$, it turns out, interestingly, the variation of the ratio $\Delta_r$, or the percentage reduction of the intensity ratio $\Delta_p$ between the 3C and 3D lines actually follows closely to a nearly universal function of the product of the reduced Debye length $\lambda_D$ and the nuclear charge $Z$, i.e., $\propto [(Z-9.2)DZ]^{-1.8}$.\ Qualitatively, this general feature results essentially from what we concluded in our study, i.e., the mixing of the $^1P_1$ component in the upper $^3D_1$ state due to the spin-dependent interactions for the 3D line is indeed responsible for the reduction of the intensity ratio $\Delta_I=gf^\mathrm{3C}(D)/gf^\mathrm{3D}(D)$ subject to the outside plasma.\ Should the general features we have identified be verified experimentally for a single Ne-like ion subject to external plasma, a simple extrapolation from a set of calculated data for this single Ne-ion could easily be extended to other Ne-like ions and, as a result, offers an alternative for plasma diagnostics.\\
\indent Finally, the general scaling behaviors of the redshifts and line ratios of the Ne-like ions found in this study are very easy for practical use. Therefore, they should be useful for the modelling and diagnostic of astrophysical plasmas such as active galactic nuclei and quasars, stellar coronae and supernovae, etc., and laboratory plasmas such as inertial and magnetic confinement fusion devices.\\
\section*{Acknowledgements}
This work is supported by the National Natural Science Foundation of China (Grant Nos. 11774023 and U1530401), the National Key Research and Development Program of China (Grant No.2016YFA0302104), the National High-Tech ICF Committee in China.\ We acknowledge the computational support provided by the Beijing Computational Science Research Center.\ In addition, T.N.C. is thankful for the partial support from the National Center for Theoretical Science (NCTS) in Taiwan.\\
%



\bibliographystyle{mnras}

\section*{References}
\begin{thebibliography}{99}
\bibitem{Tennyson2005}Tennyson J 2005 {\it Astronomical Spectroscopy: An Introduction to the Atomic and Molecular Physics of Astronomical Spectra} (London: Imperial College Press)
\bibitem{Kallman2007}Kallman T R and Palmeri P 2007 {\it \RMP} {\bf 79} 79-133
\bibitem{Lindl2004}Lindl J D, Amendt P, Berger R L, Glendinning S G, Glenzer S H, Haan S W, Kauffman R L, Landen O L  and Suter L J 2004 {\it Phys. Plasmas} {\bf 11} 339-491
\bibitem{BooZer2004}Boozer A H 2004 {\it Rev. Mod. Phys.} {\bf 76} 1071-141
\bibitem{Griem1997} Griem H R 1997 {\it Principles of Plasma Spectroscopy, Cambridge Monographs on Plasma Physics} (Cambridge: Cambridge University Press)
\bibitem{Skupsky1980}Skupsky S 1980 {\it Phys. Rev.} A {\bf 21} 1316-26
\bibitem{Behar2001}Behar E, Cottam J and Kahn S M 2001 {\bf Astrophys. J.} {\bf 548} 966-75
\bibitem{Xu2002}Xu H, Kahn S M, Peterson J R, Behar E, Paerels F B S, Mushotzky R F, Jernigan J G, Brinkman A C and Makishima K 2002 {\bf Astrophys. J.} {\bf 579} 600-6
\bibitem{Paerels2003}Paerels F B S and Kahn S M 2003 {\it Annu. Rev. Astron. Astrophys.} {\bf 41} 291-342
\bibitem{Trigo2013}D\'iaz Trigo M, Miller-Jones J C A, Migliari S, Broderick J W and Tzioumis T 2013 {\it Nature} {\bf 504} 260-2
\bibitem{Brown2001a}Brown G V, Beiersdorfer P, Chen H, Chen M H and Reed K J 2001 {\it Astrophys. J. Lett.} {\bf 557} L75-8
\bibitem{Safronova2001}Safronova U I, Namba C, Murakami I, Johnson W R and Safronova M S 2001{\it Phys. Rev.} A {\bf 64} 012507
\bibitem{Chen2002}Chen G X and Pradhan A K 2002 {\it Phys. Rev. Lett.} {\bf 89} 013202
\bibitem{Chen2007}Chen G X 2007 {\it Phys. Rev.} A {\bf 76} 062708
\bibitem{Bernitt2012}Bernitt S et al 2012 {\it Nature} {\bf 492} 225-8
\bibitem{Natalia2014}Oreshkina N S, Cavaletto S M, Keitel C H and Harman Z 2014 {\it Phys. Rev. Lett.} {\bf 113} 143001
\bibitem{Loch2015}Loch S D, Ballance C P, Li Y, Fogle M and Fontes C J 2015 {\it Astrophys. J. Lett.} {\bf 801} L13
\bibitem{Mendoza2017}Mendoza C and Bautista M A 2017 {\it Phys. Rev. Lett.} {\bf 118} 163002
\bibitem{Wang2017} Wang K, J\"onsson P, Ekman J, Brage T, Chen C Y, Fischer C F, Gaigalas G and Godefroid M 2017 {\it Phys. Rev. Lett.} {\bf 119} 189301
\bibitem{Fang2018} Fang T K, Wu C S, Gao X and Chang T N 2018 {\it Phys. Plasmas} {\bf 25} 102116
\bibitem{Nguyen1986}Nguyen H, Koenig M, Benredjem D, Caby M and Coulaud G 1986 {\it Phys. Rev.} A {\bf 33} 1279-90
\bibitem{Davis1982}Davis J and Blaha M 1982 {\it J. Quant. Spectrosc. Radiative Transfer} {\bf 27} 307-13
\bibitem{Glenzer2009}Glenzer S H and Redmer R 2009 {\it Rev. Mod. Phys.} {\bf 81} 1625-63
\bibitem{Li2017}Li X F, Jiang G, Wang H B, Wu M and Sun Q 2017 {\it Phys. Scripta} {\bf 92} 075401; Li X F, Jiang G, Wang H B and Sun Q 2017 {\it Chinease Phys.} B {\bf 26} 013101
\bibitem{Debye1923}Debye P and H\"uckel E 1923 {\it Phys. Z.} {\bf 24} 185-206
\bibitem{Chang2013}Chang T N and Fang T K 2013 {\it Phys. Rev.} A {\bf 88} 023406
\bibitem{Chang2015}Chang T N, Fang T K and Gao X 2015 {\it Phys. Rev.} A {\bf 91} 063422
\bibitem{Fang2017} Fang T K, Wu C S, Gao X and Chang T N 2017 {\it Phys. Rev.} A {\bf 96} 052502
\bibitem{Bielinska2004}Bielinska-Waz D, Karwowski J, Saha B and Mukherjee P K 2004 {\it Phys. Rev.} E {\bf 69} 016404
\bibitem{Ho_series}Kar S and Ho Y K 2005 {\it New J. Phys.} {\bf 7} 141; Kar S and Ho Y K 2006 {\it Int. J. Quantum Chemistry} {\bf 106} 814-22; Kar S and Ho Y K 2008 {\it J. Quant. Spectrosc. Radiative Transfer} {\bf 109} 445-52; Kar S and Ho Y K 2008 {\it Phys. Plasmas} {\bf 15} 013301; Ghoshal A and Ho Y K 2009 {\it \jpb} {\bf 42} 175006; Sahoo S and Ho Y K 2010 {\it J. Quant. Spectrosc. Radiative Transfer} 111 52-62; Lin C Y and Ho Y K 2010 {\it European Phys. J.} D {\bf 57} 21-6; Lin C Y and Ho Y K 2011 {\it Comput. Phys. Communications} {\bf 182} 125-9; Jiang Z S, Kar S and Ho Y K 2012 {\it Phys. Plasmas} 19 033301
\bibitem{Kar2015}Kar S and Jiang Z S 2015 {\it Atomic Data and Nuclear Data Tables} {\bf 102} 42-63
\bibitem{Saha2003}Saha B, Mukherjee P K, Bielinska D and Karwowski J 2003 {\it J. Quant. Spectrosc. Radiative Transfer} {\bf 78} 131-7
\bibitem{Deprince2017}Deprince J, Fritzsche S, Kallman T R, Palmeri P and Quinet P 2017 {\it \CJP} {\bf 95} 858-61
\bibitem{Mukherjee_series}Ray D and Mukherjee P K 1998 {\it European Phys. J.} D {\bf 2} 89-92; Mukherjee P K, Karwowski J and Diercksen G H F 2002 {\it Chemical Phys. Lett.} {\bf 363} 323-7; Sil A N and Mukherjee P K 2005 {\it Int. J. Quantum. Chemistry} {\bf 102} 1061-8; Saha J K, Bhattacharyya S, Mukherjee T K and Mukherjee P K 2010 {\it J. Quant. Spectrosc. Radiative. Transfer} {\bf 111} 675-88
\bibitem{Zhang2010} Zhang S B, Wang J G and Janev R K 2010 {\it Phys. Rev.} A {\bf 81} 032707; Zhang S B, Wang J G and Janev R K 2010 {\it Phys. Rev. Lett.} {\bf 104} 023203
\bibitem{Qi2009}Qi Y Y, Wu Y, Wang J G and Qu Y Z 2009 {\it Phys. Plasmas} {\bf 16} 023502; Qi Y Y, Wang J G and Janev R K 2016 {\it Phys. Plasmas} {\bf 23} 073302
\bibitem{Wang1995}Wang Z and Winkler P 1995 {\it Phys. Rev.} A {\bf 52} 216-20
\bibitem{Dai2001}Dai S T, Solovyova A and Winkler P 2001 {\it Phys. Rev.} E {\bf 64} 016408
\bibitem{Lopez1997}Lopez X, Sarasola C and Ugalde J M 1997 {\it J. Phys. Chemistry} A {\bf 101,} 1804-7
\bibitem{Okutsu2005}Okutsu H, Sako T, Yamanouchi K and Diercksen G H F 2005 {\it \jpb} {\bf 38} 917-27
\bibitem{Shukla2011}Shukla P K and Eliasson B 2011 {\it \RMP} {\bf 83} 885-906
\bibitem{Mondal2013}Mondal P K, Dutta N N, Dixit G and Majumder S 2013 {\it Phys. Rev.} A {\bf 87} 062502
%
\bibitem{Rouse1967}Rouse C A 1967 {\it Phys. Rev.} {\bf 163} 62-71
\bibitem{Margenau1959}Margenau H and Lewis M 1959 {\it \RMP} {\bf 31} 569-615
\bibitem{FFChen2006}Chen F F 2006 {\it Introduction to Plasma Physics and Controlled Fusion} vol~1 Plasma Physics 2nd ed (Berlin: Springer Science)
%
\bibitem{Grant2006}Grant I P 2006 {\it Relativistic Quantum Theory of Atoms and Molecules: Theory and Computation} Springer Series on Atomic Optical and Plasma Physics (New York: Springer-Verlag)
\bibitem{Jonsson2007}J\"onsson P, He X, Fischer C F and Grant I P 2007 {\it Comput. Phys. Communications} {\bf 177} 597-622
\bibitem{Han2012}Han X Y, Gao X, Zeng D L, Yan J and Li J M 2012 {\it Phys. Rev.} A {\bf 85} 062506
\bibitem{Gao2014}Gao X, Han X Y, Zeng D L, Jin R and Li J M 2014 {\it Phys. Lett.} A {\bf 378} 1514-19
\bibitem{Han2014}Han X Y, Gao X, Zeng D L, Jin R, Yan J and Li J M 2014 {\it Phys. Rev.} A {\bf 89} 042514
\bibitem{Gao2016}Gao X, Han X Y and Li J M, 2016 {\it \jpb} {\bf 49} 214005
%
\bibitem{Nantel1998}Nantel M, Ma G, Gu S, C\^ot\'e C Y, Itatani J and Umstadter D 1998 {\it Phys. Rev. Lett.} {\bf 80} 4442-5
\bibitem{Saemann1999}Saemann A, Eidmann K, Golovkin I E, Mancini R C, Andersson E, F\"orster E and Witte K 1999 {\it Phys. Rev. Lett.} {\bf 82} 4843-6
\bibitem{Leng1995}Leng Y, Goldhar J, Griem H R and Lee R W 1995 {\it Phys. Rev.} E {\bf 52} 4328-37
\bibitem{Woolsey1998}Woolsey N C et al 1998 {\it Phys. Rev.} E {\bf 57} 4650-62
\bibitem{Goldston1995}Goldston R J and Rutherford P H 1995 {\it Introduction to Plasma Physics} (London: IOP Publishing Ltd)
%
\bibitem{Santana2015}Santana J A, Lepson J K, Tr\"abert E and Beiersdorfer P 2015 {\it Phys. Rev.} A {\bf 91} 012502
\bibitem{NIST}Kramida A, Ralchenko Yu, Reader J and NIST ASD Team 2018 {\it NIST Atomic Spectra Database} (ver. 5.6.1) available
at http://physics.nist.gov/asd (Gaithersburg: National Institute of Standards and Technology)
\bibitem{Brown2001}Brown G V, Beiersdorfer P and Widmann K 2001 {\it Phys. Rev.} A {\bf 63} 032719
\bibitem{Beiersdorfer2001}Beiersdorfer P, von Goeler S, Bitter M and Thorn D B 2001 {\it Phys. Rev.} A {\bf 64} 032705
\bibitem{Bieli2004}Bieli\'nska-Wa\.z D, Karwowski J, Saha B and Mukherjee P K 2004 {\it Phys. Rev.} E {\bf 69} 016404
\bibitem{jonsson2014} J\"onsson P et al 2014 {\it Atomic Data and Nuclear Data Tables} {\bf 100} 1-154
\bibitem{Wu2019}Wu C S and Gao X to be published
\end{thebibliography}




\label{lastpage}
\end{document}